\documentclass[smus]{snow2e}
\usepackage{epsf}

\def\sgn{\mathop{\rm sgn}}
\def\etmiss{\slashchar{E}_T}

\def\Meff{M_{\rm eff}}
\def\Msusy{M_{\rm SUSY}}
\def\lsp{\tilde\chi_1^0}
\def\ra{\rightarrow}
\def\GeV{{\rm GeV}}

\let\badcite=\cite
\def\cite{~\badcite}

\def\slashchar#1{\setbox0=\hbox{$#1$}           
   \dimen0=\wd0                                 
   \setbox1=\hbox{/} \dimen1=\wd1               
   \ifdim\dimen0>\dimen1                        
      \rlap{\hbox to \dimen0{\hfil/\hfil}}      
      #1                                        
   \else                                        
      \rlap{\hbox to \dimen1{\hfil$#1$\hfil}}   
      /                                         
   \fi}                                         %

\def\dofig#1#2{\centerline{\epsfxsize=#1\epsfbox{#2}}}

\font\twelvess=cmss10 scaled \magstep1

\begin{document}

\begingroup
\parindent=20pt
\vbox{\vskip12mm}
\moveright7mm\vbox to 8.9in{\hsize=6.5in
{\large
\centerline{\twelvess BROOKHAVEN NATIONAL LABORATORY}
\vskip6pt
\hrule
\vskip1pt
\hrule
\vskip4pt
\hbox to \hsize{August, 1996 \hfil BNL--\phantom{00000}}
\vskip3pt
\hrule
\vskip1pt
\hrule
\vskip3pt

\vskip1.5in
\centerline{\LARGE\bf Determining SUSY Particle Masses at LHC}
\vskip.5in
\centerline{\bf Frank E. Paige}
\vskip4pt
\centerline{Physics Department}
\centerline{Brookhaven National Laboratory}
\centerline{Upton, NY 11973 USA}
\vskip1in
\centerline{ABSTRACT}
\vskip8pt
Some possible methods to determine at the LHC masses of SUSY particles
are discussed.
\vskip1in
	To appear in the {\sl Proceedings of the Workshop on Future
Directions in High Energy Physics} (Snowmass, 1996).
}

\vfil

	This manuscript has been authored under contract number
DE-AC02-76CH00016 with the U.S. Department of Energy.  Accordingly,
the U.S.  Government retains a non-exclusive, royalty-free license to
publish or reproduce the published form of this contribution, or allow
others to do so, for U.S. Government purposes.
}

\vfill\eject\endgroup

\title{Determining SUSY Particle Masses at LHC}

\author{Frank E. Paige\\ {\it Physics Department, Brookhaven National
Laboratory, Upton, NY 11973, USA}} 

\maketitle

\begin{abstract}
Some possible methods to determine at the LHC masses of SUSY particles
are discussed.
\end{abstract}

\thispagestyle{empty}\pagestyle{empty}

\section{Introduction}

	If supersymmetry (SUSY) exists at the electroweak scale, it
should be easy at the LHC to observe deviations from the Standard
Model (SM) such as an excess of events with multiple jets plus missing
energy $\etmiss$ or with like-sign dileptons $\ell^\pm\ell^\pm$ plus
$\etmiss$\cite{ATLAS,CMS,BCPT}. Determining SUSY masses is more
difficult because each SUSY event contains two missing lightest SUSY
particles $\lsp$, and there are not enough kinematic constraints to
determine the momenta of these. This note describes two possible
approaches to determining SUSY masses, one based on a generic global
variable and the other based on constructing particular decay chains.

	The ATLAS and CMS Collaborations at the LHC are considering
five points in the minimal supergravity (SUGRA) model listed in
Table~I below\cite{LHC}. Point~4 is the comparison point extensively
discussed elsewhere in these Proceedings. For this point a good
strategy at the LHC is to use the decays $\tilde\chi_2^0 \ra
\lsp \ell^+\ell^-$ to determine the mass difference
$M(\tilde\chi_2^0) - M(\lsp)$\cite{LHC}. For higher masses,
e.g.{} Points~1--3, this decay is small, but $\tilde\chi_2^0 \ra
\lsp h \ra \lsp b \bar b$, $\tilde\chi_2^\pm \ra
\lsp W^\pm \ra \lsp q \bar q$, and $\tilde\chi_2^0
\ra \tilde\ell \ell \ra \lsp \ell\ell$ provide alternative
starting points for detailed analysis.

\begin{table}[ht]
\caption{SUGRA parameters for the five LHC points.}
\begin{center}
\begin{tabular}{cccccc}
\hline\hline
Point & $m_0$ & $m_{1/2}$ & $A_0$ & $\tan\beta$ & $\sgn{\mu}$ \\
      & (GeV) & (GeV)   & (GeV)   &             &             \\
\hline
1 & 100 & 300 & 300 & \phantom{0}2.1 & $+$\\
2 & 400 & 400 &   0 & \phantom{0}2.0 & $+$\\
3 & 400 & 400 &   0 & 10.0 & $+$\\
4 & 200 & 100 &   0 & \phantom{0}2.0 & $-$\\
5 & 800 & 200 &   0 & 10.0 & $+$ \\
\hline\hline
\end{tabular}
\end{center}
\vskip-4pt
\end{table}

\begin{figure}[ht]
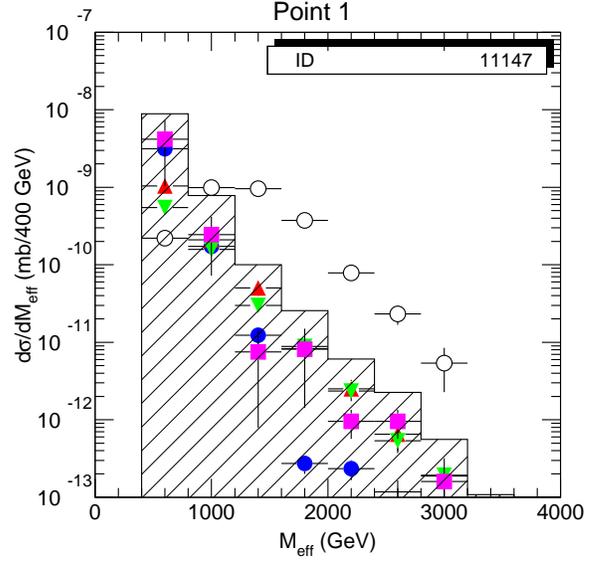

\dofig{2.95in}{point1_147.epsi}
\caption{Point~1 signal and backgrounds. Open circles: signal.  Solid
circles: $t\bar t$. Triangles: $W\ra\ell\nu$, $\tau\nu$.  Downward
triangles:  $Z\ra\nu\bar\nu$, $\tau\tau$. Squares: QCD jets.
Histogram: all backgrounds.}
\end{figure}

\begin{figure}[ht]
\dofig{2.95in}{point2_147.epsi}
\caption{Signal and SM backgrounds for Point~2. See Fig.~1 for symbols.}
\end{figure}

\begin{figure}[t]
\dofig{2.95in}{point3_147.epsi}
\caption{Signal and SM backgrounds for Point~3. See Fig.~1 for symbols.}
\end{figure}

\begin{figure}[ht]
\dofig{2.95in}{point4_147.epsi}
\caption{Signal and SM backgrounds for Point~4. See Fig.~1 for symbols.}
\end{figure}

\begin{figure}[ht]
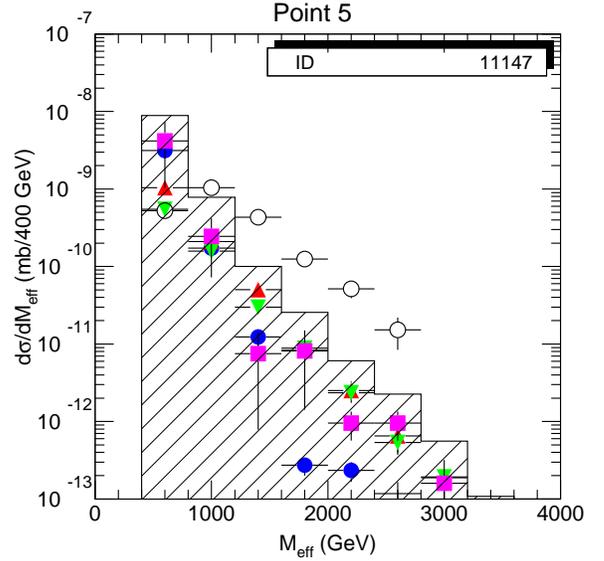

\dofig{2.95in}{point5_147.epsi}
\caption{Signal and SM backgrounds for Point~5. See Fig.~1 for symbols.}
\end{figure}

\section{Effective Mass Analysis}

	The first step after discovering a deviation from the SM is to
estimate the mass scale. SUSY production at the LHC is dominated by
gluinos and squarks, which decay into jets plus missing energy. The
mass scale can be estimated using the effective mass, defined as the 
scalar sum of the $p_T$'s of the four hardest jets and the missing
transverse energy $\etmiss$,
$$
\Meff = p_{T,1} + p_{T,2} + p_{T,3} + p_{T,4} + \etmiss\,.
$$

	ISAJET~7.20\cite{ISAJET} was used to generate samples of 10K
events for each signal point, 50K events for each of $t \bar t$, $Wj$
with $W \to e\nu,\mu\nu,\tau\nu$, and $Zj$ with $Z \to
\nu\bar\nu,\tau\tau$ in five bins covering $50 < p_T < 1600\,\GeV$,
and 2500K QCD events, i.e., primary $g$, $u$, $d$, $s$, $c$, or $b$
jets, in five bins covering $50 < p_T < 2400\,\GeV$. The detector
response was simulated using a toy calorimeter with
\begin{eqnarray}
{\rm EMCAL} &\quad& 10\%/\sqrt{E} + 1\% \nonumber\\
{\rm HCAL}  &\quad& 50\%/\sqrt{E} + 3\% \nonumber\\
{\rm FCAL}  &\quad& 100\%/\sqrt{E} + 7\%,\ |\eta| > 3\,.\nonumber
\end{eqnarray}
Jets were found using a simple fixed-cone algorithm (GETJET) with
$R=[(\Delta\eta)^2+(\Delta\phi)^2]^{1/2}=0.7$. To suppress the SM
background, the following cuts were made:  
\begin{itemize}
\item	$\etmiss > 100\,\GeV$
\item	$\ge4$ jets with $p_T > 50\,\GeV$ and  $p_{T,1} > 100\,\GeV$
\item	Transverse sphericity $S_T > 0.2$
\item	Lepton veto
\item	$\etmiss > 0.2 \Meff$
\end{itemize}
With these cuts and the idealized detector assumed here, the signal is
much larger than the SM backgrounds for large $\Meff$, as is
illustrated in Figs.~1--5. 

\begin{table}[b]
\caption{The value of $\Meff$ for which $S = B$ compared to $\Msusy$,
the lighter of the gluino and squark masses. Note that Point~4 is
strongly influenced by the $\etmiss$ and jet $p_T$ cuts.}
\begin{center}
\begin{tabular}{cccc}
\hline\hline
Point& $\Meff\,(\GeV)$& $M_{\rm SUSY}\,(\GeV)$& Ratio\\
\hline
1 &      \phantom{0}980 &   663 &    1.48 \\
2 &      1360           &   926 &    1.47 \\
3 &      1420           &   928 &    1.53 \\
4 &      \phantom{0}470 &   300 &    1.58 \\
5 &      \phantom{0}980 &   586 &    1.67 \\
\hline\hline
\end{tabular}
\vskip-10pt
\end{center}
\end{table}

	The peak of the $\Meff$ mass distribution, or alternatively
the point at which the signal and background are equal, provides a
good first estimate of the SUSY mass scale, which is defined to be
$$
\Msusy = \min(M_{\tilde g}, M_{\tilde u_R})
$$ 
(The choice of $M_{\tilde u_R}$ as the typical squark mass is
arbitrary.)  The ratio of the value $\Meff$ for which $S = B$ to
$\Msusy$ was calculated by fitting smooth curves to the signal and
background and is given in Table~II. To see whether the approximate
constancy of this ratio might be an accident, 100 SUGRA models were
chosen at random with $100 < m_0 < 500\,\GeV$, $100 < m_{1/2} <
500\,\GeV$, $-500 < A_0 < 500\,\GeV$, $1.8 < \tan\beta < 12$, and
$\sgn\mu=\pm1$ and compared to the assumed signal, Point~1. The light
Higgs was assumed to be known, and all the comparison models were
required to have $M_h = 100.4 \pm 3\,\GeV$. A sample of 1K events was
generated for each point, and the peak of the $\Meff$ distribution was
found by fitting a Gaussian near the peak.  Figure~6 shows the
resulting scatter plot of $\Msusy$ vs.{} $\Meff$.  The ratio is
constant within about $\pm10\%$, as can be seen from Fig.~7. This
error is conservative, since there considerable contribution to the
scatter from the limited statistics and the rather crude manner in
which the peak was found.

\begin{figure}[t]
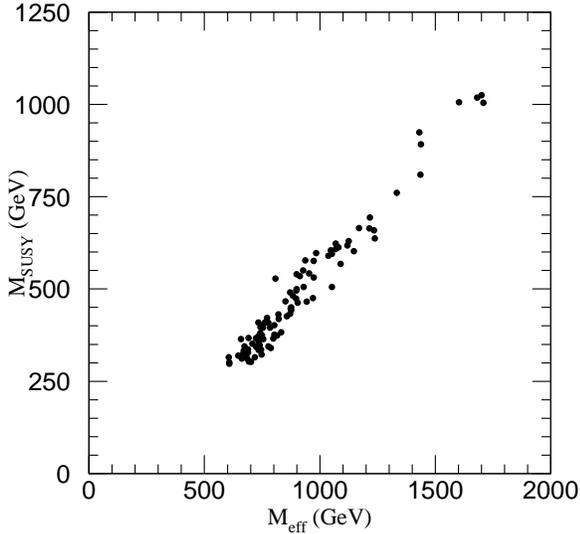

\dofig{2.95in}{scan1.epsi}
\caption{Scatter plot of $\Msusy = \min(M_{\tilde g}, M_{\tilde u})$
vs.{} $\Meff$ for randomly chosen SUGRA models having the same light
Higgs mass within $\pm3\,\GeV$ as Point~1.}
\end{figure}

\begin{figure}[t]
\dofig{2.95in}{scan3.epsi}
\caption{Ratio $\Meff/\Msusy$ from Fig.~6}
\end{figure}

\def\idiots{$h \ra b \bar b$}
\section{Selection of \idiots}

\begin{figure}[ht]
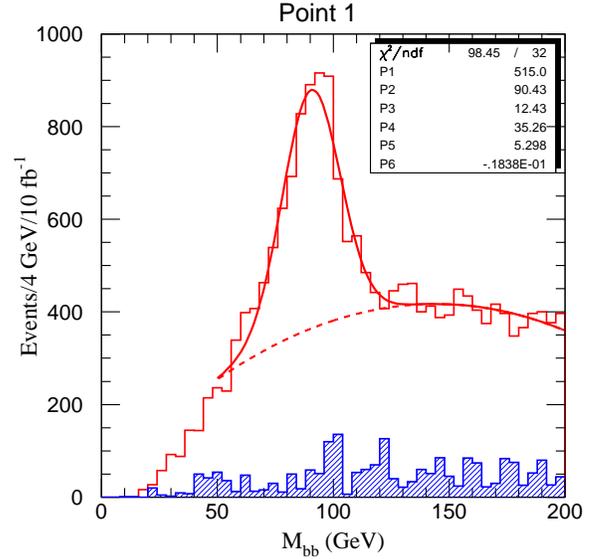

\dofig{2.95in}{x1_mbb.epsi}
\caption{$M(b \bar b)$ for pairs of $b$ jets for the Point~1 signal
(open histogram) and for the sum of all backgrounds (shaded histogram)
after cuts described in the text. The smooth curve is a Gaussian plus
quadratic fit to the signal. The light Higgs mass is $100.4\,\GeV$.}
\end{figure}

	For Point~1 the decay chain $\tilde\chi_2^0 \ra \lsp
h$, $h \ra b \bar b$ has a large branching ratio, as is typical if
this decay is kinematically allowed. The decay $h \ra b \bar b$ thus
provides a handle for identifying events containing
$\tilde\chi_2^0$'s\cite{ERW}. Furthermore, the gluino is heavier than
the squarks and so decays into them. The strategy for this analysis is
to select events in which one squark decays via
$$
\tilde q \ra \tilde\chi_2^0 q,\ \tilde\chi_2^0 \ra \lsp h,\
h \ra b \bar b\,,
$$
and the other via
$$
\tilde q \ra \lsp q\,,
$$
giving two $b$ jets and exactly two additional hard jets.

	ISAJET~7.22\cite{ISAJET} was used to generate a sample of 100K
events for Point~1, corresponding to about $5.6\,{\rm fb}^{-1}$.
Background samples of 250K each for $t \bar t$, $Wj$, and $Zj$, and
5000K for QCD jets were also generated, equally divided among five
$p_T$ bins. The background samples generally represent a small
fraction of an LHC year. The detector response was simulated using the
toy calorimeter described above.  Jets were found using a fixed cone
algorithm with $R = 0.4$. The following cuts were imposed:
\begin{itemize}
\item	$\etmiss > 100\,\GeV$
\item	$\ge4$ jets with $p_T > 50\,\GeV$ and  $p_{T,1} > 100\,\GeV$
\item	Transverse sphericity $S_T > 0.2$
\item	$\Meff > 800\,\GeV$
\item	$\etmiss > 0.2 \Meff$
\end{itemize}
Jets were tagged as $b$'s if they contained a $B$ hadron with $p_T >
5\,\GeV$ and $\eta < 2$; no other tagging inefficiency or $b$
mistagging was included. Figure~8 shows the resulting $b \bar b$ mass
distributions for the signal and the sum of all SM backgrounds with
$p_{T,b} > 25\,\GeV$ together with a Gaussian plus quadratic fit to
the signal.  At a luminosity of $10^{33}\,{\rm cm}^{-2} {\rm s}^{-1}$,
\hbox{ATLAS} will have a $b$-jet tagging efficiency of 70\% for a
rejection of 100\cite{ATLAS}.  Hence, the number of events should be
reduced by a factor of about two, but the mistagging background is
probably small compared to the real background shown. The Higgs mass
peak is shifted downward somewhat; using a larger cone, $R = 0.7$,
gives a peak which is closer to the true mass but wider.

\begin{figure}[t]
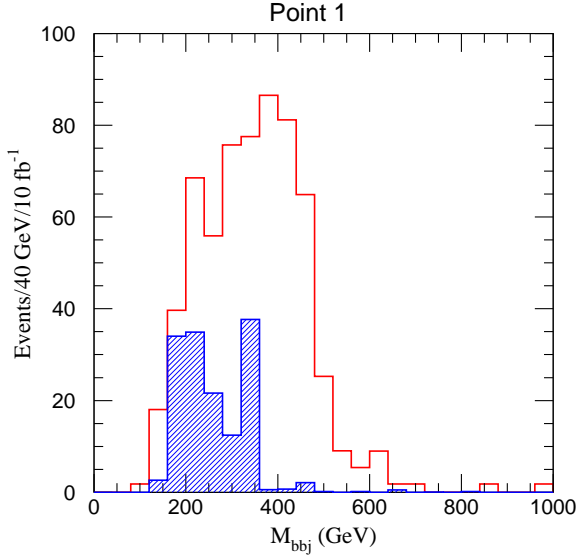

\dofig{2.95in}{x1_mbbj.epsi}
\caption{The smaller of the two $b \bar b j$ masses for signal and
background events with $73 < M(b \bar b) < 111\,\GeV$ in Fig.~8 and
with exactly two additional jets $j$ with $p_T > 75\,\GeV$. The
endpoint of this distribution should be approximately the mass
difference between the squark and the $\tilde \chi_1^0$, about
$542\,\GeV$.} 
\end{figure}

	Events were then required to have exactly one $b \bar b$ pair
with $73 < M(b \bar b) < 111\,\GeV$ and exactly two additional jets
with $p_T > 75\,\GeV$. The invariant mass of each jet with the $b \bar
b$ pair was calculated.  For the desired decay chain, one of these two
must come from the decay of a single squark, so the smaller of them
must be less than the kinematic limit for single squark decay,
$M(\tilde u_R) -M(\lsp) = 542\,\GeV$.  The smaller of the two $b \bar
b j$ masses is plotted in Fig.~9 for the signal and for the sum of all
backgrounds and shows the expected edge. The SM background shows
fluctuations from the limited Monte Carlo statistics but seems to be
small near the edge, at least for the idealized detector considered
here. There is some background from the SUSY events above the edge,
presumably from other decay modes and/or initial state radiation. 

\def\idiots{$W \ra q \bar q$}
\section{Selection of \idiots}

\begin{figure}[b]
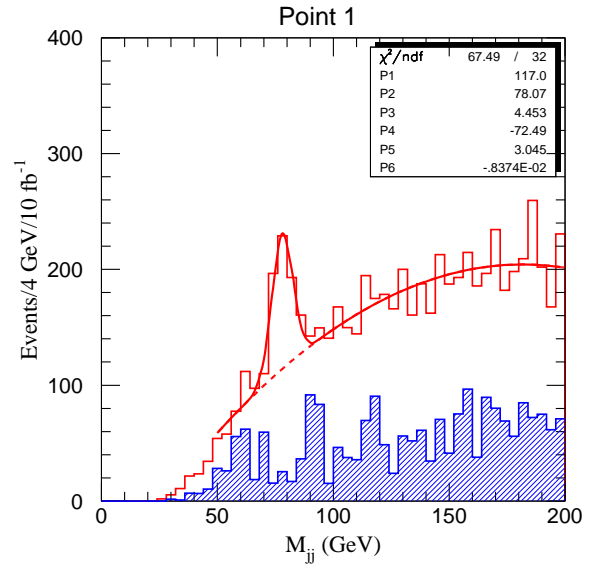

\dofig{2.95in}{x1_mjj.epsi}
\caption{$M_{34}$ for non-$b$ jets in events with two $200\,\GeV$
jets and two $50\,\GeV$ jets for the Point~1 signal (open histogram)
and the sum of all backgrounds (shaded histogram).}
\end{figure}

	Point~1 also has a large combined branching ratio for one
gluino to decay via
$$
\tilde g \ra \tilde q_L \bar q,\ \tilde q_L \ra \tilde\chi_1^\pm q,\
\tilde\chi_1^\pm \ra \lsp W^\pm,\ 
W^\pm \ra q \bar q\,,
$$
and the other via
$$
\tilde g \ra \tilde q_R q,\ \tilde q_R \ra \lsp q\,,
$$
giving two hard jets and two softer jets from the $W$. The branching
ratio for $\tilde q_L \ra \lsp q$ is small for Point~1, so
the contributions from $\tilde g \ra \tilde q_L \bar q$ and from
$\tilde q_L \tilde q_L$ pair production are suppressed.

	The same signal sample was used as in Section~III, and jets
were again found using a fixed cone algorithm with $R = 0.4$. The
combinatorial background for this decay chain is much larger than for
the previous one, so harder cuts are needed:
\begin{itemize}
\item	$\etmiss > 100\,\GeV$
\item	$\ge4$ jets with $p_{T1,2} > 200\,\GeV$, $p_{T3,4} > 
50\,\GeV$, and \hfil $\eta_{3,4} < 2$
\item	Transverse sphericity $S_T > 0.2$
\item	$\Meff > 800\,\GeV$
\item	$\etmiss > 0.2 \Meff$
\end{itemize} 
The same $b$-tagging algorithm was applied to tag the third and fourth
jets as not being $b$ jets. Of course, this is not really feasible;
instead one should measure the $b$-jet distributions and subtract
them.  

	The mass distribution $M_{34}$ of the third and fourth highest
$p_T$ jets with these cuts is shown in Fig.~10 for the signal and the
sum of all backgrounds. A peak is seen a bit below the $W$ mass with a
fitted width surprisingly smaller than that for the $h$ in Fig.~8,
note that the $W$ natural width has been neglected in the simulation
of the decays. The SM background is more significant here than for $h
\to b \bar b$. Events from this peak can be combined with another jet
as was done for $h \ra b \bar b$ in Fig.~9, providing another
determination of the squark mass. Figure~10 also provides a starting
point for measuring $W$ decays separately from other sources of
leptons such as gaugino decays into sleptons.
\vfill

\def\idiots{$\tilde\chi_2^0 \ra \tilde\ell \ell \ra \lsp \ell\ell$}
\section{Selection of \idiots}

\begin{figure}[b]
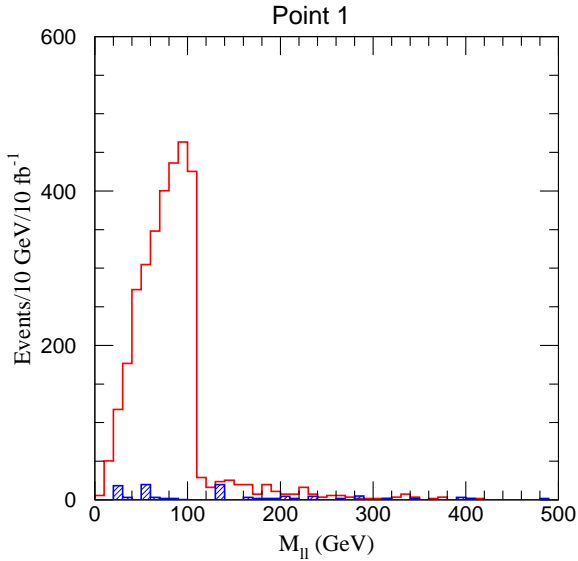

\dofig{2.95in}{x1_mll.epsi}
\caption{$M_{\ell\ell}$ for the Point~1 signal (open histogram) and
the sum of all backgrounds (shaded histogram).}
\end{figure}

\begin{figure}[ht]
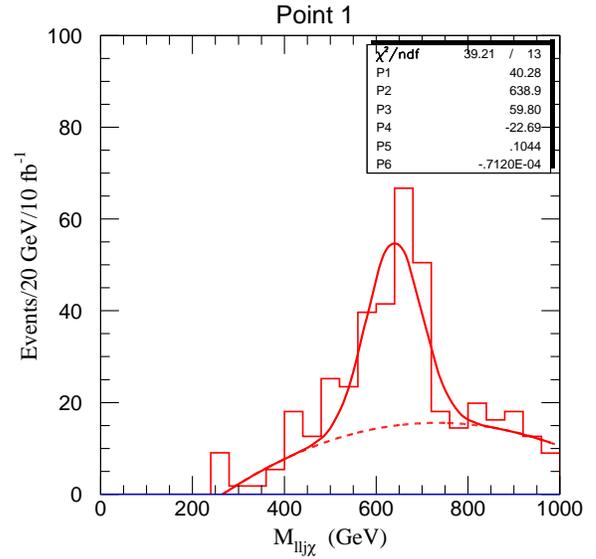

\dofig{2.95in}{x1_mlljh.epsi}
\caption{$M_{\ell\ell j\lsp}$ for events with $86 < M_{\ell\ell} <
109\,\GeV$ in Fig.~11, using $\vec p_{\lsp} = M_{\lsp} / M_{\ell\ell}
\vec p_{\ell\ell}$ for the Point~1 signal (open histogram) and the
SM background (shaded histogram).}
\end{figure}

	Point~1 has relatively light sleptons, which is generically
necessary if the $\lsp$ is to provide acceptable cold dark
matter\cite{BB}. Hence 
the two-body decay
$$
\tilde\chi_2^0 \ra \tilde\ell_R \ell \ra \lsp \ell^+\ell^-
$$
is kinematically allowed and competes with the $\tilde\chi_2^0 \ra
\lsp h$ decay, producing opposite-sign, like-flavor dileptons. The
largest SM background is $t \bar t$. To suppress this and other SM
backgrounds the following cuts were made on the same signal and SM
background samples used in the two previous sections:
\begin{itemize}
\item	$\Meff > 800\,\GeV$ 
\item	$\etmiss > 0.2\Meff$
\item	$\ge 1$ $R=0.4$ jet with $p_{T,1} > 100\,\GeV$
\item	$\ell^+\ell^-$ pair with $p_{T,\ell}> 10\,\GeV$, $\eta_\ell < 2.5$
\item	$\ell$ isolation cut: $E_T < 10\,\GeV$ in $R=0.2$
\item	Transverse sphericity $S_T > 0.2$
\end{itemize}
With these cuts very little SM background survives, and the
$M_{\ell\ell}$ mass distribution shown in Fig.~11 has an edge near
$$
M_{\ell\ell}^{\rm max} = M_{\tilde\chi_2^0} 
\sqrt{1-{M_{\tilde\ell}^2 \over M_{\tilde\chi_2^0}^2}}
\sqrt{1-{M_{\lsp}^2 \over M_{\tilde\ell}^2}} \approx 112\,\GeV\,,
$$

	If $M_{\ell\ell}$ is near its kinematic limit, then the
velocity difference of the $\ell^+\ell^-$ pair and the $\lsp$ is
minimized. Having both leptons hard requires $M_{\tilde\ell}/
M_{\tilde\chi_2^0}^2 \sim M_{\lsp} / M_{\tilde\ell}$. Assuming this
and $M_{\tilde\chi_2^0} = 2 M_{\lsp}$ implies that the endpoint in
Fig.~11 is equal to the $\lsp$ mass. An improved estimate could be made
by detailed fitting of all the kinematic distributions. Events were
selected with $M_{\ell\ell}^{\rm max} -10\,\GeV < M_{\ell\ell} <
M_{\ell\ell}^{\rm max}$, and the $\lsp$ momentum was calculated using
this crude $\lsp$ mass and
$$
\vec p_{\lsp} = (M_{\lsp} / M_{\ell\ell})\,\vec p_{\ell\ell}\,.
$$
The invariant mass $M_{\ell\ell j\lsp}$ of the $\ell^+\ell^-$, the
highest $p_T$ jet, and the $\lsp$ was then calculated and is shown in
Fig.~12.  A peak is seen near the light squark masses,
660--$688\,\GeV$. More study is needed, but this approach looks
promising.

\vfil

	This work would have been impossible without the contributions
of my collaborators on ISAJET, H. Baer, S. Protopopescu, and X. Tata.
It was supported in part by the United States Department of Energy
under contract DE-AC02-76CH00016.

\vfilneg
\end{document}